\documentclass[12pt]{article}
\usepackage[hidelinks]{hyperref}
\usepackage{cite}
\usepackage{xcolor}
\usepackage{graphicx}
\usepackage{caption,subcaption}
\usepackage{amsmath}
\usepackage{amssymb}
\usepackage{xspace}
\usepackage{verbatim}
\usepackage{mathtools}
\usepackage{tikz-cd}
\usepackage{braket}
\usepackage{url}
\usepackage[normalem]{ulem}

\makeatletter
\@addtoreset{equation}{section}

\makeatletter
\renewcommand\section{\@startsection {section}{1}{\z@}%
                                   {-3.5ex \@plus -1ex \@minus -.2ex}%nn
                                   {2.3ex \@plus.2ex}%
                                   {\normalfont\large\bfseries}}
\renewcommand\subsection{\@startsection{subsection}{2}{\z@}%
                                     {-3.25ex\@plus -1ex \@minus -.2ex}%
                                     {1.5ex \@plus .2ex}%
                                     {\normalfont\bfseries}}

\def\baselinestretch{1.2}
\newcommand{\be}{\begin{equation}}
\newcommand{\ee}{\end{equation}}
\newcommand{\beq}{\begin{eqnarray}}
\newcommand{\eeq}{\end{eqnarray}}

\newcommand{\gone}[1]{{}}

\newcommand{\re}{{\rm Re}}
\newcommand{\im}{{\rm Im}}
\newcommand{\nn}{\nonumber}

\begin{document}
\begin{titlepage}
\begin{flushright}
MAD-TH-20-01
\end{flushright}

\vfil

\begin{center}

{\bf \Large
Chaos and complementarity in de Sitter space
}

\vfil

Lars Aalsma and Gary Shiu

\vfil

laalsma@wisc.edu, shiu@physics.wisc.edu

Department of Physics, University of Wisconsin, Madison, WI 53706, USA

\vfil
\end{center}

%%%%%%%%%%%%%%%%%%%%%%
\begin{abstract}

\noindent We consider small perturbations to a static three-dimensional de Sitter geometry. For early enough perturbations that satisfy the null energy condition, the result is a shockwave geometry that leads to a time advance in the trajectory of geodesics crossing it. This brings the opposite poles of de Sitter space into causal contact with each other, much like a traversable wormhole in Anti-de Sitter space. In this background, we compute out-of-time-order correlators (OTOCs) to asses the chaotic nature of the de Sitter horizon and find that it is maximally chaotic: one of the OTOCs we study decays exponentially with a Lyapunov exponent that saturates the chaos bound. We discuss the consequences of our results for de Sitter complementarity and inflation.

\end{abstract}
\vspace{0.5in}

\end{titlepage}
\renewcommand{\baselinestretch}{1.05} 

\tableofcontents

\section{Introduction}
Over the last years, it has been realized that quantum chaos plays an important role in the physics of black holes. The key property that makes black holes chaotic is the large blueshift between an asymptotic and a freely falling observer. Any perturbation with a small energy $E_0$ experiences a boost in energy given by $ E =  E_0e^{\frac{2\pi}\beta t}$, where $t$ is the Killing time used by an asymptotic observer and $\beta$ is the inverse temperature of the black hole. One probe of chaos in quantum systems that already has been known for a long time is the double commutator of two generic operators $V,W$ \cite{Larkin1969}
\be
C(t) = \braket{-[V(0),W(t)]^2} ~,
\ee
which measures the sensitivity of the operators $W$ and $V$ with respect to each other. For Hermitian and unitary operators $V$ and $W$, we can write
\be
C(t) = 2-2\,\re{\braket{V(0)W(t)V(0)W(t)}} ~,
\ee
where
\be \label{eq:introOTOC}
F(t) = \braket{V(0)W(t)V(0)W(t)} ~,
\ee
is referred to as the out-of-time-order correlator (OTOC). Chaotic behaviour shows itself in an exponential growth of the double commutator $C(t)$ or, equivalently, an exponential decay of the OTOC $F(t)$. In some thermal systems with a large number of degrees of freedom $N$, such as holographic CFTs dual to black holes, $F(t)$ behaves as \cite{Shenker:2013pqa,Shenker:2014cwa,Roberts:2014ifa,Maldacena:2015waa}
\be \label{eq:holoOTOC}
F(t) = 1- \frac{f_0}{N}e^{\lambda_Lt} + {\cal O}(N^{-2})~, \qquad \left(\beta/2\pi \ll t\ll \lambda_L^{-1}\log(N)\right) ~,
\ee
such that $C(t)\sim N^{-1}e^{\lambda_L t}$. Here $f_0$ is a positive order one constant. The timescale when $F(t)$ is affected by an order one amount is known as the scrambling time $t_* = \lambda_L^{-1}\log(N)$ and $\lambda_L$ as the (quantum) Lyapunov exponent. The size of the Lyapunov exponent determines how fast chaos can grow and it has been argued that it obeys the universal bound \cite{Maldacena:2015waa}: $\lambda_L \leq 2\pi/\beta$. Famously, black holes saturate this bound making them among  the fastest scrambling systems in nature \cite{Sekino:2008he}. Any perturbation to a black hole `scrambles' as fast as possible over the horizon, making it indistinguishable from its thermal atmosphere.

Because these developments have offered a window into the microscopic description of black holes, one might hope to similarly apply some of these tools to cosmological spacetimes. In fact, a black hole horizon shares similarities with the cosmological horizon of the static patch of de Sitter space. For instance, there is a large blueshift between an observer sitting at center of the static patch and one that is freely falling through the horizon of the first observer. Just as for black hole spacetimes, when a perturbation is released a scrambling time ($t_*=\frac{\beta}{2\pi}\log(S)$) to the past of the $t=0$ slice or earlier than that, the boosted perturbation creates a high-energy shockwave. This observation has led Susskind to conjecture that de Sitter space is also a fast scrambler \cite{Susskind:2011ap}. From this perspective, it seems natural that de Sitter space should also be maximally chaotic, i.e. it should saturate the chaos bound.

However, there are also important differences between cosmological and black hole horizons. One of the most important difference in this context is the fact that shockwaves generated by matter that obeys the null energy condition (NEC) have different properties in de Sitter space than in Minkowski or Anti-de Sitter space. Whereas geodesics crossing a positive-energy shockwave experience a gravitational time delay in Minkowski and Anti-de Sitter space, they enjoy a time advance in de Sitter space \cite{Gao:2000ga}. In this sense, a perturbation to de Sitter space that obeys the NEC has similar properties as a traversable wormhole in Anti-de Sitter space \cite{Gao:2016bin,Maldacena:2017axo}, because it now becomes possible to send signals from otherwise causally disconnected regions.

Another difference of de Sitter space as compared to black holes in Anti-de Sitter space is the absence of a spatially asymptotic and non-gravitating boundary theory from which we can probe the static patch. The only boundaries in de Sitter space are spacelike and have access to a larger region than just a single static patch. Therefore, to study chaos we restrict ourselves to a single static observer which spontaneously breaks the isometry group of $d$-dimensional de Sitter space from $SO(d,1)\to SO(d-1)\times {\mathbb R}$. This perspective has previously been taken in \cite{Parikh:2004wh,Anninos:2011af} to study a putative holographic dual of the de Sitter static patch and in \cite{Aalsma:2019rpt} to explore vacuum state modifications.

The main aim of this paper is to compute OTOCs in the static patch of de Sitter space to study chaos. Although there have been previous studies of chaos and quantum information in de Sitter space (such as the recent papers \cite{Anninos:2018svg,Geng:2019bnn,Geng:2019ruz,Bhattacharyya:2020rpy}), to the best of our knowledge no direct computation of OTOCs in de Sitter space has ever been published that demonstrates chaotic behaviour. Our goal is to fill this gap. In order to do so we find it convenient to work in $2+1$ dimensions, making the computation rather tractable. This allows us to calculate various OTOCs with operators inserted at the origin of different static patches and establish that a particular single-sided OTOC exhibits Lyapunov behaviour: it decays with a Lyapunov exponent that saturates the chaos bound. Interestingly, we find that the OTOC does not decay precisely as in \eqref{eq:holoOTOC}, but behaves as $F(t) \sim 1- N^{-2} e^{2\lambda_Lt}$ leading to $C(t)\sim N^{-2}e^{2\lambda_L t}$. This behaviour of the double commutator is the same as for black holes in Einstein gravity, but in that case this is caused by the fact that the leading term in the OTOC is purely imaginary \cite{Shenker:2014cwa}. However, we find that in de Sitter space the leading term in the OTOC is real. This seems to be an important distinction between chaos in black holes and de Sitter space.

This article is organized as follows. In section \ref{sec:dSbasics} we remind the reader about some of the basics of de Sitter space and discuss coordinate systems, Wightman functions and shockwave geometries. Section \ref{sec:OTOCs} contains the main results of our paper, where we compute various OTOCs in de Sitter space. Finally, we discuss the implications of our results for de Sitter complementarity and inflation in section \ref{sec:complementarity} and end with a discussion in section \ref{sec:discussion}.

\section{Basics of de Sitter space} \label{sec:dSbasics}

\subsection{Coordinate systems}
\begin{figure}[h]
\centering
\includegraphics[scale=.5]{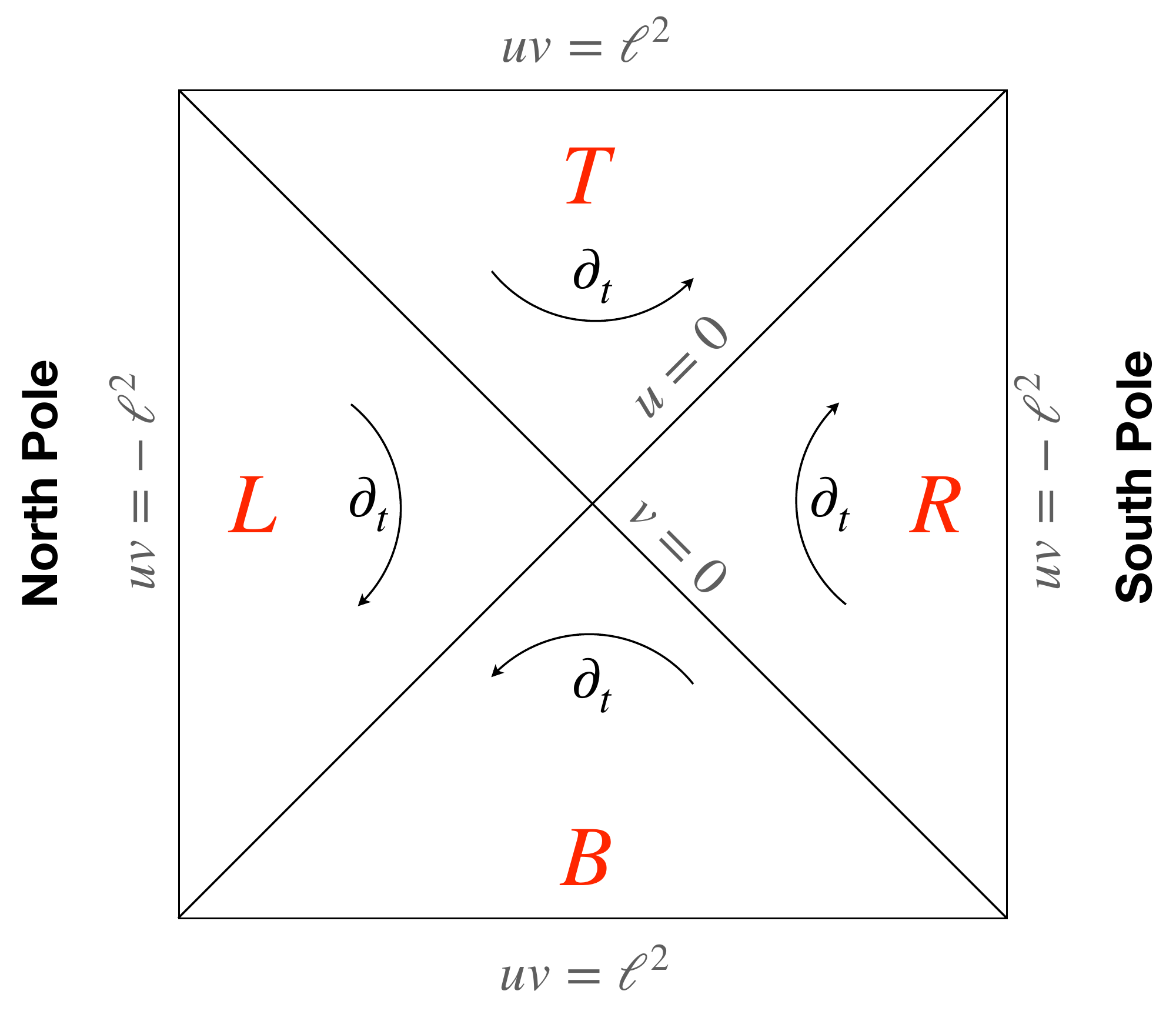}
\caption{Penrose diagram of de Sitter space. By complexifying the static time coordinate, we can cover each of the four static patches. The flow of the timelike Killing vector $\partial_t$ is indicated with arrows in each patch.}
\label{fig:PenrosedS}
\end{figure}

De Sitter space in $d$ dimensions can be described as a hyperboloid embedded into $d+1$ dimensional Minkowski space using embedding coordinates $X^{A=0,d}$:
\be
\eta_{AB}X^AX^B=\ell^2 ~.
\ee
Here $\ell$ is the de Sitter length and $\eta_{AB}$ is the Minkowski metric. A useful coordinate system in which time translation invariance is manifest are the so-called static coordinates.
\begin{align}
X^0 &= \sqrt{\ell^2-r^2}\sinh(t/\ell) ~,\\
X^d &= \sqrt{\ell^2-r^2}\cosh(t/\ell) ~, \nonumber\\
X^i &= r y^i ~. \nonumber
\end{align}
Here $y^{i=1,d-1}$ are coordinates on the unit $d-2$ sphere. The metric in this coordinate system is given by
\be \label{eq:staticmetric}
ds^2 = -\left(1-r^2/\ell^2\right)dt^2 + \left(1-r^2/\ell^2\right)^{-1}dr^2 + r^2d\Omega_{d-2}^2 ~.
\ee
This metric only covers a quarter of the global de Sitter Penrose diagram known as the static patch, surrounded by a horizon at $r=\ell$. It will be convenient to complexify the static time coordinate by writing $t_x = t + i \epsilon_x$. We can then cover any of the four static patches of the Penrose diagram, which we refer to as the right ($R$), left ($L$), top ($T$), and bottom ($B$) patch by considering different imaginary parts as follows.
\be
\epsilon_R=0 ~, \quad \epsilon_L = -\pi \ell ~, \quad \epsilon_T = -\frac\pi2\ell ~, \quad \epsilon_B = \frac\pi2\ell ~.
\ee
The Penrose diagram is displayed in figure \ref{fig:PenrosedS}.
Another metric that we will use that provides a global cover of de Sitter space is given by the coordinates
\begin{align} \label{eq:KruskalEmbed}
X^0 &= \frac{\ell^2(u+v)}{\ell^2-uv} ~, \\
X^d &= \frac{\ell^2(u-v)}{\ell^2-uv} ~, \nonumber \\
X^i &= \frac{\ell^2+uv}{\ell^2-uv} \ell y^i ~. \nonumber
\end{align}
The metric in this coordinate system is given by
\be
ds^2 = \frac{4\ell^4}{(\ell^2-uv)^2}(-dudv) + \ell^2\frac{(\ell^2+uv)^2}{(\ell^2-uv)^2}d\Omega_{d-2}^2 ~.
\ee
In this coordinate system, the past horizon is given by $v=0$ and the future horizon by $u=0$. The North and South pole are given by $uv=-\ell^2$ and the future and past boundaries by $uv=\ell^2$.

\subsection{Wightman function}
We can define a particular vacuum state $\ket{\Omega}$ by considering the Wightman function ${\cal W}(x,y)$. It is given by the two-point function of scalar fields.
\be
{\cal W}(x,y) \equiv \bra{\Omega}\varphi(x)\varphi(y) \ket{\Omega} ~.
\ee
Here $\varphi$ is a massive scalar field described by the action
\be
S = -\frac1{2}\int d^dx\sqrt{-g}\left(\partial_\mu\varphi\partial^\mu\varphi + m^2\varphi^2 +\xi R\varphi^2\right) ~,
\ee
with $\xi$ a non-minimal coupling. For states that preserve all de Sitter isometries, the Wightman function can only depend on the de Sitter invariant distance
\be
Z(x,y) = \frac1{\ell^{2}}\eta_{AB}X^A(x)X^B(y) ~.
\ee
In the Bunch-Davies vacuum the Wightman function is given by (see for example \cite{Einhorn:2002nu})
\be \label{eq:WigthmanBD}
{\cal W}(x,y) = \frac{\Gamma(h_+)\Gamma(h_-)}{\ell^d(4\pi)^{d/2}\Gamma(d/2)} \,_2F_1\left(h_+,h_-,\frac d2;\frac{1+Z(x,y)}{2}\right) ~.
\ee
Here,
\be
h_\pm = \frac12\left(d-1\pm\sqrt{(d-1)^2-4\ell^2\tilde m^2}\right) ~,
\ee
with $\tilde m^2 = m^2  + \xi R$. It is important to notice that the parameters $h_\pm$ are only purely real for masses $\tilde m^2\ell^2 \leq (d-1)^2/4$.  The distinction between the real and imaginary regimes of $h_\pm$ can be made in terms of representations of the isometry group of de Sitter space, $SO(d,1)$. The range of masses $0<\tilde m^2\ell^2<\frac{(d-1)^2}{4}$ corresponds to the complementary series representation and $\tilde m^2\ell^2 \geq \frac{(d-1)^2}{4}$ to the principal series representation \cite{Newton1941,Newton1950}.

The Wightman function \eqref{eq:WigthmanBD} is analytic everywhere in the complex $Z$ plane except at a branch cut along the line $Z\geq1$. For timelike separated points $Z>1$ we therefore need to regularize the Wightman function and the correct $i\epsilon$ prescription is to send $Z(x,y)\to Z(x,y) + i\epsilon\,\text{sgn}(x,y)$ \cite{Einhorn:2002nu}.\footnote{The sign difference of our prescription with respect to \cite{Einhorn:2002nu} comes from the different choice of metric signature.} We define $\text{sgn}(x,y)$ to be $+1$ when $x$ is to the future of $y$ and $-1$ when $x$ is in the past of $y$. Thus, the properly regularized Wightman function in the Bunch-Davies vacuum is given by
\be
{\cal W}(x,y) =  \frac{\Gamma(h_+)\Gamma(h_-)}{\ell^d(4\pi)^{d/2}\Gamma(d/2)} \,_2F_1\left(h_+,h_-,\frac d2;\frac{1+Z(x,y)+i\epsilon\,\text{sgn}(x,y)}{2}\right) ~.
\ee

\subsection{Shockwaves}
Let us now focus on the $R$ patch. The relation between static and global coordinates is given by
\be
u = - \ell e^{-t/\ell} \sqrt{\frac{\ell-r}{\ell+r}} ~, \quad v = \ell e^{t/\ell} \sqrt{\frac{\ell-r}{\ell+r}} ~.
\ee
We then see that a time translation $t\to t + c$ corresponds to a boost in global coordinates.
\be
u \to  e^{-c/\ell}u ~, \quad v \to e^{c/\ell}v ~.
\ee
This shows that a particle released from the origin of the static patch a time $t$ to the past of the $t=0$ slice will be highly blueshifted when it crosses the $t=0$ slice. It is therefore appropriate to describe such a particle as a shockwave geometry.

\begin{figure}[h]
\centering
\includegraphics[scale=.25]{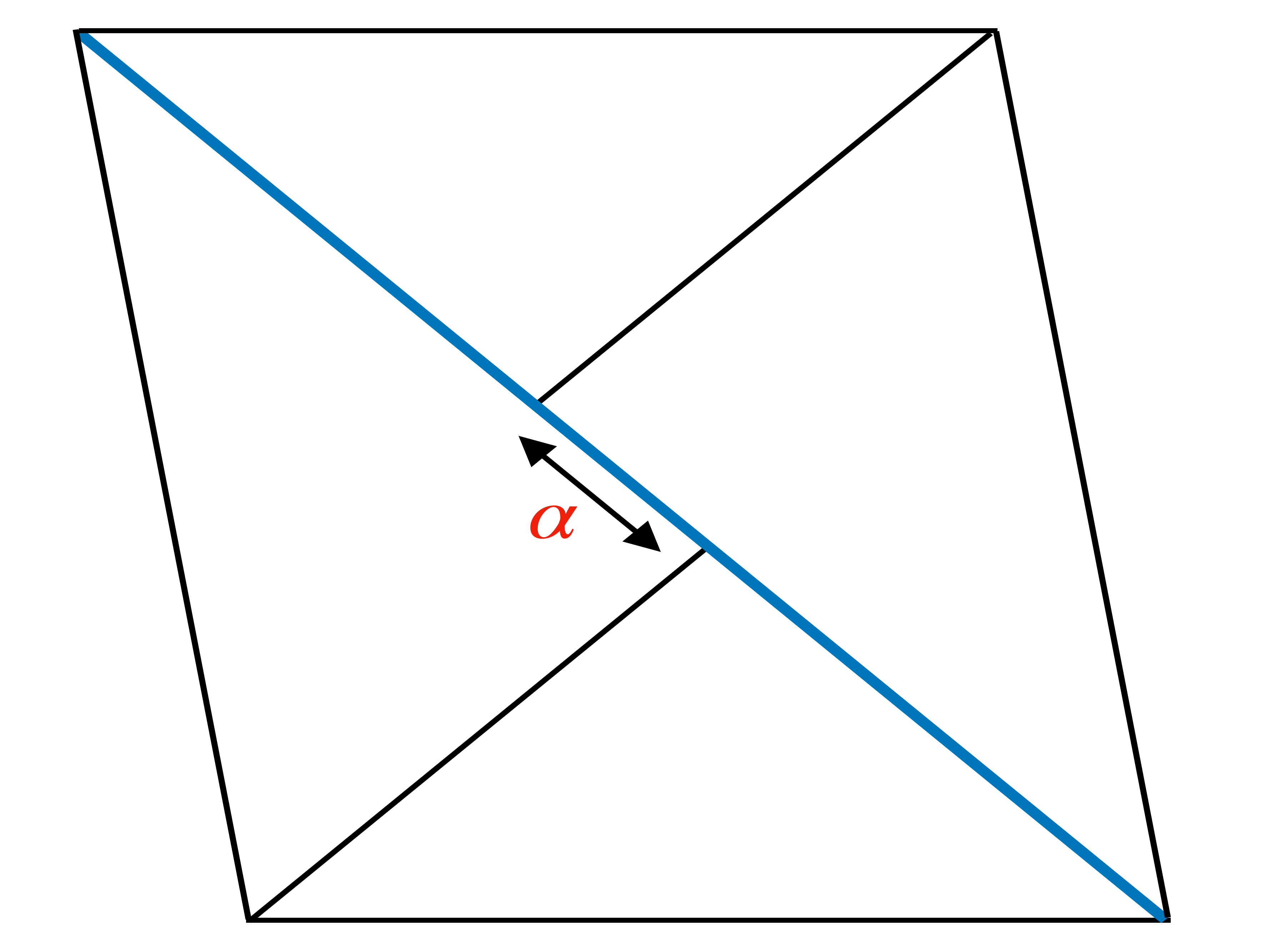}
\caption{Penrose diagram of the shockwave geometry \eqref{eq:discontinuousshock} created by a highly boosted particle that travels along the past horizon $v=0$ (the blue line). There is a discontinuity in the coordinate $\tilde u$ by an amount $\alpha$ which brings the left and right static patch into causal contact with each other.}
\label{fig:penshock}
\end{figure}

We will focus on $2+1$ dimensions, but higher-dimensional de Sitter shockwave geometries have also been constructed, see for example \cite{Hotta:1992qy,Hotta:1993,Sfetsos:1994xa}. For shockwaves travelling along the past horizon $v=0$, the metric is given by (see appendix \ref{app:shockwave})
\be \label{eq:continuousshock}
ds^2 = \frac{4\ell^4}{(\ell^2-uv)^2}(-dudv) -4\alpha\delta(v)dv^2 + \ell^2\left(\frac{\ell^2+uv}{\ell^2-uv}\right)^2d\phi^2 ~.
\ee
Here $\phi=\phi+2\pi$ and we ignored the spread of the shockwave in the transverse direction for now. Geodesics crossing the past horizon $v=0$ in this metric experience a time advance by an amount $\alpha$. This is a solution to Einstein's equations with a stress tensor given by
\be
T_{vv} = \frac{\alpha}{4\pi G_N\ell^2}\delta(v) ~.
\ee
The null energy condition enforces $\alpha>0$. If this shockwave is generated by a particle which in its restframe has a thermal energy given by $E_0=\beta^{-1}= (2\pi \ell)^{-1}$ the parameter $\alpha$ is related to the blueshifted energy by (see appendix \ref{app:shockwave})
\be
\alpha = G_N e^{t_w/\ell} ~.
\ee
Here $t_w=-t$ is the time the particle is released to the past of the $t=0$ slice. The Penrose diagram of this geometry is shown in figure \ref{fig:penshock}.
It will sometimes be convenient to also consider the metric in a slightly different form by performing the coordinate transformation $u = \tilde u-\alpha\theta(v)$. We then find
\be \label{eq:discontinuousshock}
ds^2 = \frac{4\ell^4}{(\ell^2-(\tilde u-\alpha\theta(v))v)^2}(-d\tilde udv) + \ell^2\left(\frac{\ell^2+(\tilde u-\alpha\theta(v))v}{\ell^2-(\tilde u-\alpha\theta(v))v}\right)^2d\phi^2 ~,
\ee
Here $\theta(v)$ is the Heaviside theta function. In this metric, there is a discontinuity in the $\tilde u$ coordinate at $v=0$ by an amount $\alpha$. Shockwaves with positive null energy can therefore bring opposite poles of de Sitter space into causal contact with each other \cite{Gao:2000ga}. 

\section{Out-of-time-order correlators} \label{sec:OTOCs}
OTOCs are specified by their analytical continuation from Euclidean four-point functions, but they still have a freedom that corresponds to moving the operators along the thermal circle, see e.g. figure 2 of \cite{Shenker:2014cwa}. These different configurations are all related by analytical continuation, but as we will see this does not mean that they all show chaotic behaviour. To assess chaos, we will focus on two different configurations. The first configuration that we study is obtained by moving one of the operators halfway along the thermal circle. This corresponds to the double-sided correlator $\braket{W_RV_LV_RW_R}$. The second configuration is the purely single-sided configuration $\braket{W_RV_RV_RW_R}$. The subscripts here refer to either the left or right static patch of the de Sitter Penrose diagram, as indicated in figure \ref{fig:PenrosedS}. We will now compute the different OTOCs.

\subsection{Geodesic approximation}
First, we will calculate the following OTOC that was previously considered in the context of black holes by Shenker and Stanford \cite{Shenker:2013pqa}.
\be \label{eq:geodOTOC}
F(t) = \braket{W_R(t)V_L(0)V_R(0)W_R(t)} ~,
\ee
We will evaluate this correlation function in the Bunch-Davies state. The operators $W_R$ and $V_{L,R}$ correspond to massive scalar fields inserted at the origin of a static patch indicated by the subscript. Alternatively, we can see it as a purely right-sided correlator where we evaluate the operator with subscript $L$ at time $-i\pi\ell$ to move it to the left side. Notice that this particular ordering is only equal to $\braket{V_L(0)W_R(t)V_R(0)W_R(t)}$ when $V_L$ and $W_R$ are spacelike separated. To calculate this correlation function, we will make use of a geodesic approximation. We can view $F(t)$ as a two-point function in the shockwave background which is given by a path integral over all possible paths connecting the two operators. For earlier work exploiting the geodesic approximation in the context of AdS/CFT, see \cite{Balasubramanian:1999zv,Louko:2000tp,Kraus:2002iv,Fidkowski:2003nf,Kaplan:2004qe,Festuccia:2005pi}. In the limit of large mass $m\ell \gg 1$ of the $V$ operators, the path integral is solved by a saddle point approximation in which the two-point function localizes to a sum over geodesics with the location of the operators as the end points:
\be \label{eq:geodapprox}
F(t) \simeq \sum_{\rm geodesics} e^{-m D} ~.
\ee
$D$ is the (renormalized) geodesic distance, which in a de Sitter background is given by
\be
\cos\left(\frac{D(x,y)}\ell\right) = Z(x,y) ~.
\ee
We should proceed with some caution, because \eqref{eq:geodapprox} is only unambiguous for operators in a geometry with a real analytic continuation. In that case, the geodesic distance can straightforwardly be computed in Euclidean signature and the Lorentzian correlator is obtained by analytical continuation. However, we are in a situation where this condition is not true since the shockwave induces some non-analyticity in the metric. Nonetheless, seeing the shockwave as a small perturbation to the background geometry we expect that \eqref{eq:geodapprox} still gives the dominant contribution, just as in \cite{Shenker:2013pqa}. A more careful treatment would be to introduce an auxiliary spacetime that has a real analytical continuation and a limit in which it reduces to the Lorentzian shockwave geometry, such as was done in \cite{Balasubramanian:2012tu}.

\begin{figure}[h]
\centering
\includegraphics[scale=.5]{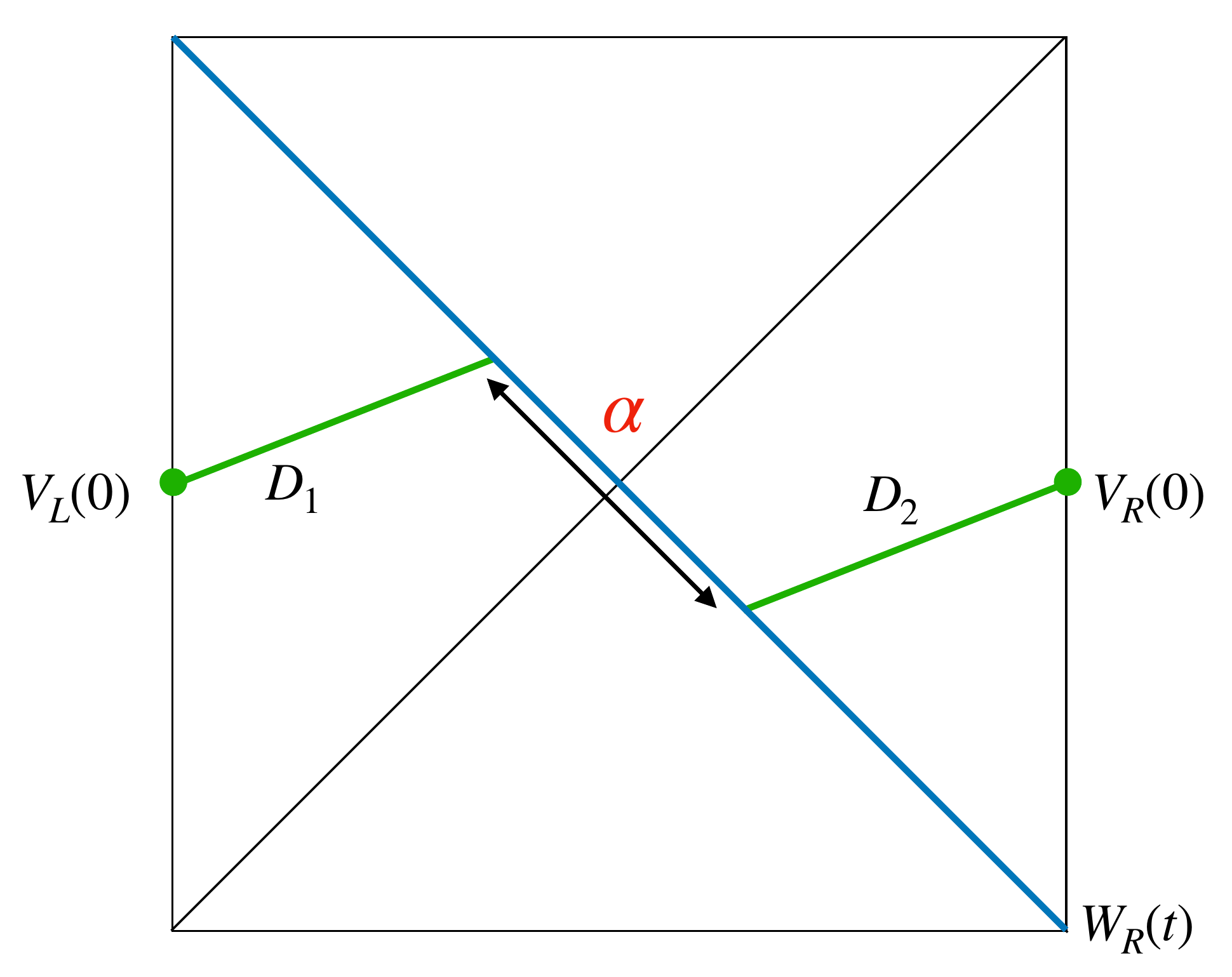}
\caption{The geodesic (green) connecting the operators $V_L(0)$ and $V_R(0)$ in the shockwave geometry \eqref{eq:continuousshock} created by the operator $W_R(t)$. Due to the shockwave (blue) the geodesic is shifted by an amount $\alpha$ along the past horizon.}
\label{fig:geodapprox}
\end{figure}

\begin{figure}[h]
\centering
\includegraphics[scale=.75]{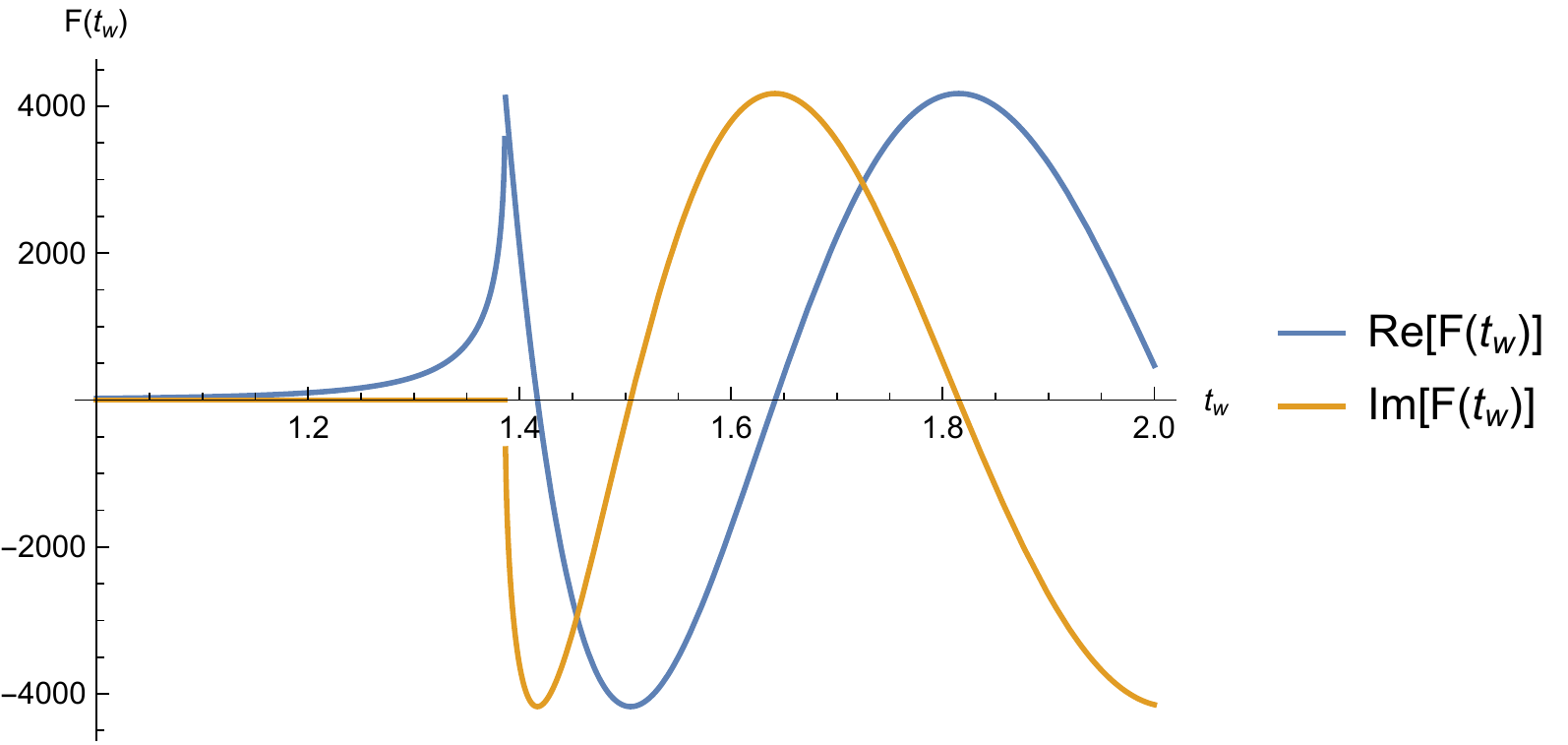}
\caption{The OTOC $F(t_w)$ calculated using the geodesic approximation which is valid for $m^2\ell^2\gg 1$. It develops oscillations after the scrambling time $t>t_*=\ell \log(S_{dS})$. To plot the figure we took $G_N = 1/2, \ell = 1$ and $m^2\ell^2=10$. }
\label{fig:OTOC_geod}
\end{figure}

Putting this subtlety aside for now, we will proceed to calculate the geodesic distance between the $V_L$ operators in two parts. Using the embedding coordinates \eqref{eq:KruskalEmbed}, we first calculate the distance $D_1$ between $V_L(0)$ and the shockwave at $v=0$. Then, we add to it $D_2$: the distance from the horizon $v=0$ to the operator $V_R$, see figure \ref{fig:geodapprox}.
We find
\be
\cos\left(\frac{D_1}\ell\right) = \frac{u}{\ell} ~, \quad \cos\left(\frac{D_2}\ell\right) = \frac{\alpha-u}{\ell} ~.
\ee
Thus, the total geodesic distance is given by
\be
D= D_1+D_2 = \ell\arccos\left(\frac u\ell\right) + \ell\arccos\left(\frac {\alpha-u}\ell\right) ~.
\ee
Extremizing over $u$, we find that $u=\alpha/2$, which leads to
\be
D = 2\ell\arccos\left(\frac{\alpha}{2\ell}\right) ~.
\ee
This results in a correlation function given by
\be
F(\alpha) = e^{-2m\ell \arccos\left(\frac{\alpha}{2\ell}\right)} ~.
\ee
Expanding for $\alpha \ll 2\ell$, normalizing, and writing the result as a function of $t_w$ we find
\be \label{eq:geodestimate}
F(t_w) = 1+ mG_Ne^{t_w/\ell} + {\cal O}\left(\frac{G_N}{\ell}e^{t_w/\ell}\right)^2~.
\ee
This expansion is valid for times
\be
t_w \ll \ell \log\left(\frac{2\ell}{G_N}\right) ~.
\ee
We recognize this as the scrambling time $t_w \ll t_*=\ell\log(S_{dS})$ up to a constant that is subdominant when $S_{dS}\gg 1$. Here $S_{dS}=\pi\ell/2G_N$ is the de Sitter entropy. Notice that unlike the OTOC in black hole backgrounds \eqref{eq:geodestimate} does not decay, but grows exponentially. This is not unexpected, since we know that the effect of a positive energy shockwave is to causally connect the $L$ and $R$ patches. This can be seen from the geodesic distance. The operators  $V_L$ and $V_R$ are only spacelike separated when $\alpha<2\ell$, become null separated at $\alpha=2\ell$, and timelike when $\alpha>2\ell$.  For timelike separation ($\alpha>2\ell$), the correlation function picks up an imaginary part and starts to oscillate.
\begin{align} \label{eq:lateOTOC}
\re{(F(t_w))} &= +\cos\left(2m\ell \left|\arccos\left(\frac{G_N}{2\ell}e^{t_w/\ell}\right)\right|\right) ~,\\
\im{(F(t_w))} &= -\sin\left(2m\ell \left|\arccos\left(\frac{G_N}{2\ell}e^{t_w/\ell}\right)\right|\right) ~.\nonumber
\end{align}
We show the behaviour of the complete OTOC in figure \ref{fig:OTOC_geod}.
So clearly, the particular OTOC \eqref{eq:geodOTOC} which displayed chaotic behaviour (exponential decay) in a black hole background does not do so in de Sitter space. As we just mentioned, this should not come as a complete surprise due to the different nature of positive energy shockwaves in de Sitter space. The oscillating behaviour after the scrambling also seems to be explained by this, because already the Wightman function in a pure de Sitter background oscillates for masses $\tilde m\ell \gg 1$, which is the regime where the geodesic approximation is valid. Such heavy fields correspond to the principal series representation of $SO(3,1)$. In contrast, the Wightman function for light fields $\tilde m\ell < 1$ that fall into the complementary series representation does not exhibit oscillations and we expect qualitatively different behaviour of the OTOC in that case. In the next section, we will study the OTOC in more detail by going beyond the geodesic approximation and focussing on conformally coupled scalar fields $\tilde m^2\ell^2 = 3/4$. As we will see then, the oscillations present in the OTOC for heavy fields are indeed absent and the fact that the OTOC picks up an imaginary part has a nice interpretation in terms of information being exchanged between the left and right static patch. Moreover, we find that the purely single-sided OTOC does display Lyapunov behaviour.

\subsection{Beyond the geodesic approximation} \label{sec:scattering}
Another way of computing the OTOC was put forward by Shenker and Stanford in \cite{Shenker:2014cwa}. We skip the full derivation here and simply highlight the main ingredients going into the derivation. We do so for completeness, so that we can later compare this with our results of the OTOC in de Sitter space. In \cite{Shenker:2014cwa}, the four-point function was viewed
as the overlap between an `in' state and  `out' state created by perturbing the thermofield double state with the operators $V,W$. These states are then given by
\be
\ket{\Psi} = V_R(t_3)W_L(t_4) \ket{{\rm TFD}} ~, \quad \ket{\Psi'} = W_R(t_2)^\dagger V_L(t_1)^\dagger \ket{{\rm TFD}} ~.
\ee
For large time separation $|t_2-t_1|$ there is a large relative boost between the energies of the $W$ and $V$ particles. This implies that in an appropriate frame the $W$ particle can be viewed as a shockwave travelling close to the horizon and computing the overlap between $\ket{\Psi}$ and $\ket{\Psi'}$ becomes a high-energy scattering problem. We can now represent the `in' and `out' states as Klein-Gordon wave functions which are represented in terms of longitudinal momentum and transverse  separation. In an elastic Eikonal approximation, the full overlap is simply given by the overlap of the wave functions weighted by the Eikonal phase $e^{i\delta(s,|x-x'|)}$, which is a function of the center-of-mass energy $s=4p_1^up_2^v$ and transverse separation $|x-x'|$.

Following this procedure, the final result for the four-point correlation function is then given by \cite{Shenker:2014cwa}
\begin{align} \label{eq:gencorr}
&\braket{V_{x_1}(t_1)W_{x_2}(t_2)V_{x_3}(t_3)W_{x_4}(t_4)}  \\
&= \frac{16}{\pi^2}\int{\cal D} e^{i\delta(s,|x-x'|)}\left[p_1^u\psi_1^*(p_1^u,x)\psi_3(p_1^u,x)\right]\left[p_2^v\psi_2^*(p_2^v,x')\psi_4(p_2^v,x')\right] \nonumber ~.
\end{align}
The measure in this integral is given by ${\cal D}=\ell dx dx' dp_1^udp_2^v$ and the wave functions $\psi_i$ are given by
\begin{align} \label{eq:fouriertransforms}
\psi_1(p^u,x) &= \int dv e^{2ip^uv} \left.\braket{V(u,v,x)V_{x_1}(t_1)^\dagger}\right|_{u=0} ~, \\
\psi_2(p^v,x) &= \int du e^{2ip^vu} \left.\braket{W(u,v,x)W_{x_2}(t_2)^\dagger}\right|_{v=0} ~, \nonumber \\
\psi_3(p^u,x) &= \int dv e^{2ip^uv} \left.\braket{V(u,v,x)V_{x_3}(t_3)}\right|_{u=0} ~, \nonumber \\
\psi_4(p^v,x) &= \int du e^{2ip^vu} \left.\braket{W(u,v,x)W_{x_4}(t_4)}\right|_{v=0} ~. \nonumber 
\end{align}
The expressions derived in \cite{Shenker:2014cwa} are appropriate for planar black holes (although similar expressions have been derived for black holes with different topologies, see e.g. \cite{Ahn:2019rnq,Balasubramanian:2019stt,Balasubramanian:2019qwk}) in asymptotically Anti-de Sitter spacetimes. To apply \eqref{eq:gencorr} to de Sitter space we need to make some modifications. In the Anti-de Sitter case, the expectation values appearing in the wave functions are bulk-to-boundary propagators. Since we are interested in studying scattering of particles that are released from the center of a static patch of de Sitter space we have to replace these expectation values by Wightman functions in the Bunch-Davies vacuum. Furthermore, the transverse direction in our case is a compact circle instead of a line. Thus, the integration measure is now given by ${\cal D} =\ell^3 d\phi d\phi' dp_1^udp_2^v$.

\begin{figure}[h]
\centering
\includegraphics[scale=.8]{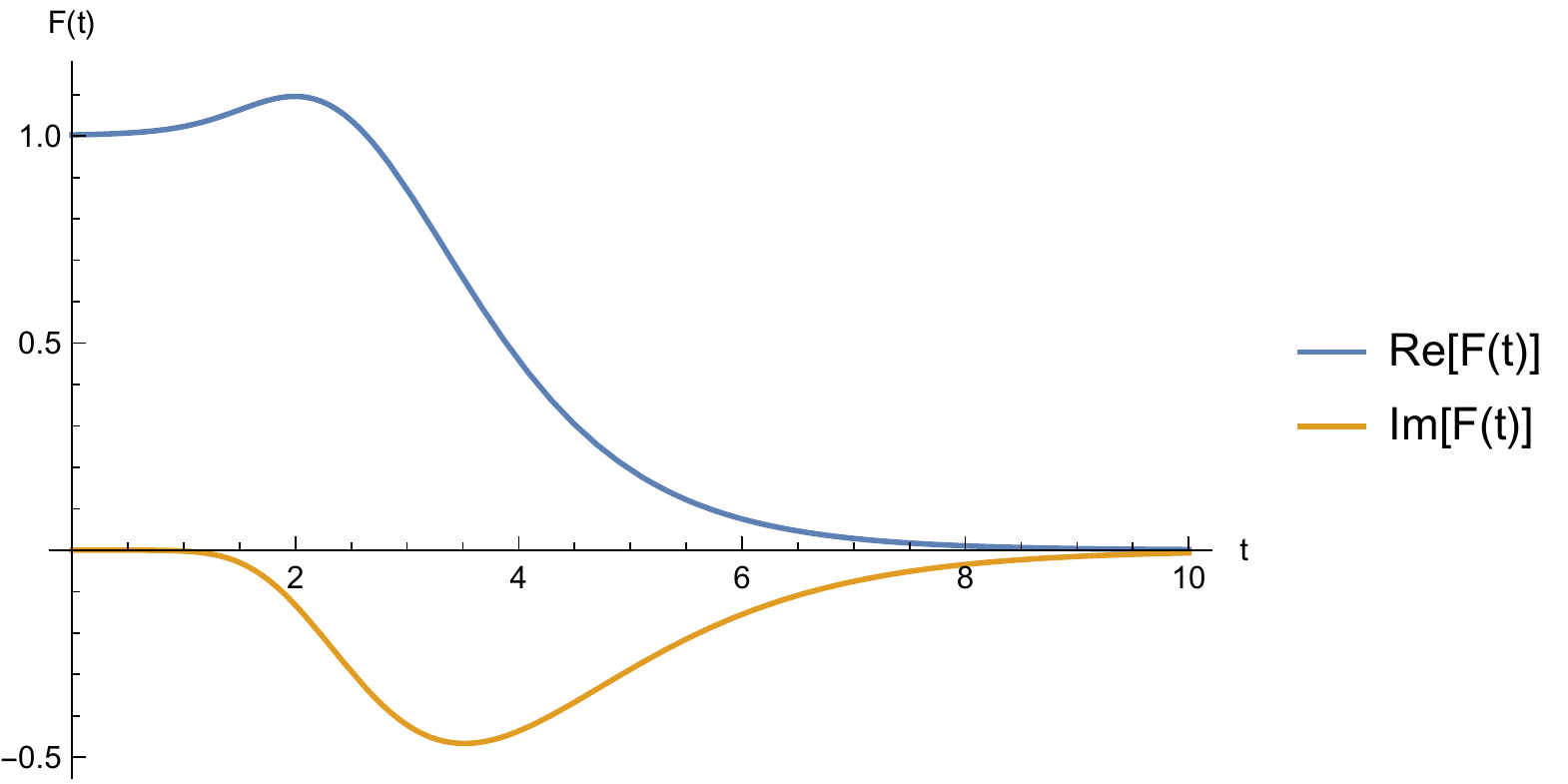}
\caption{The out-of-time-order correlator $\braket{V_L(0)W_R(t)V_R(0)W_R(t)}$ for $|g(t=0)|=10,\ell=1$.}
\label{fig:OTOC}
\end{figure}

Now that we have spelled out the main differences, we compute the wave functions in de Sitter space by taking the Fourier transform of the Wightman function. Unfortunately, because of the rather complicated form of the Wightman function in terms of a hypergeometric function, it is not easy to evaluate the integrals \eqref{eq:fouriertransforms} analytically for arbitrary masses. Instead, we will consider the more tractable situation in which all particles are conformally coupled, i.e. $\tilde m^2\ell^2 = 3/4$. The Wightman function then greatly simplifies to
\be
{\cal W}(x,y) = \frac{1}{4\sqrt{2}\ell^3\pi}\frac1{\sqrt{1-Z(x,y)-i\epsilon\,\text{sgn}(x,y)}} ~.
\ee
We can now explicitly perform the Fourier transforms to find
\begin{align} \label{eq:wavefunction}
\psi_1(p^u) &= \frac{c}{\sqrt{4\pi \ell^5 p^u}} \exp\left(2i\ell p^u e^{t_1^*/\ell} + \frac{t_1^*}{2\ell}\right) ~,\\
\psi_2(p^v) &= \frac{-c}{\sqrt{4\pi \ell^5 p^v}} \exp\left(-2i\ell p^ve^{-t_2^*/\ell}-\frac{t_2^*}{2\ell}\right) ~,\nonumber \\
\psi_3(p^u) &= \frac{c}{\sqrt{4\pi \ell^5 p^u}}\exp\left(2i\ell p^u e^{t_3/\ell} + \frac{t_3}{2\ell}\right) ~,\nonumber\\
\psi_4(p^v) &= \frac{-c}{\sqrt{4\pi\ell^5 p^v}} \exp\left(-2i\ell p^v e^{-t_4/\ell} - \frac{t_4}{2\ell}\right) \nonumber ~.
\end{align} 
Here $c$ is an unimportant constant that obeys $|c|^2=1$. Notice that the dependence on the transverse direction has dropped. We are now interested in computing the correlation function
\be
F(t) = \frac{\braket{V(i\epsilon_1)W(t+i\epsilon_2)V(i\epsilon_3)W(t+i\epsilon_4)}}{\braket{V(i\epsilon_1)V(i\epsilon_3)} \braket{W(i\epsilon_2)W(i\epsilon_4)}} ~.
\ee
The denominator of this expression is given by the general formula \eqref{eq:gencorr}, with the Eikonal phase $\delta$ set to zero. Plugging the expressions for the wave functions \eqref{eq:wavefunction} into the general formula \eqref{eq:gencorr} we find
\begin{align} \label{eq:OTOCint}
&\braket{V(i\epsilon_1)W(t+i\epsilon_2)V(i\epsilon_3)W(t+i\epsilon_4)} = \\
&\qquad \frac1{16\pi^4\ell^{10}}e^{\Delta_{12}+\Delta_{34}}\int{\cal D} e^{i\delta}  \exp\left(-2\ell p^u\epsilon_{13} - 2\ell e^{-t/\ell} p^v\epsilon_{24}^* - t/\ell\right) ~. \nonumber
\end{align}
Here we introduced the notation
\begin{align}
\epsilon_{ij} &= i \left(e^{i\epsilon_i/\ell} - e^{i\epsilon_j/\ell}\right) ~, \\
\Delta_{ij} &= \frac{i}{2\ell}(\epsilon_i-\epsilon_j) \nonumber ~.
\end{align}
To evaluate this integral, the last piece of information we need is the Eikonal phase $\delta$, which is given by the classical action \cite{Shenker:2014cwa,Balasubramanian:2019qwk}
\be
\delta = \frac12 \int d^3x\sqrt{-g}\left[\frac1{16\pi G_N}h_{uu}{\cal D}^2h_{vv} + h_{uu}T^{uu} + h_{vv}T^{vv}\right] ~.
\ee
Here $h_{uu},h_{vv}$ are the metric components corresponding to a perturbation to a pure de Sitter geometry by two shockwaves travelling along the future and past horizon. $T_{uu},T_{vv}$ are the corresponding stress tensor components. To evaluate this integral, we can use the expressions from appendix \ref{app:shockwave}. The stress tensor and metric components that solve the linear Einstein equations are given by
\begin{align}
h_{uu} &= -8\pi G_Np^v\ell\delta(u)|\sin(\phi-\phi')| ~, \qquad T_{uu} = \frac{p^v}{2\ell}\delta(u)\delta(\sin(\phi-\phi'))  ~,  \\
h_{vv} &=-8\pi G_Np^u\ell\delta(v)|\sin(\phi-\phi'')| ~, \qquad T_{vv} = \frac{p^u}{2\ell}\delta(v)\delta(\sin(\phi-\phi'')) ~. \nonumber
\end{align}
Using these expressions we find that the Eikonal phase is given by
\be
\delta =-\pi G_N\ell p^up^v |\sin(\phi'-\phi'')| ~,
\ee
Plugging this into \eqref{eq:OTOCint} we obtain
\begin{align}
&\braket{V(i\epsilon_1)W(t+i\epsilon_2)V(i\epsilon_3)W(t+i\epsilon_4)} = \\
&\qquad  \frac1{16\pi^4\ell^{10}}e^{\Delta_{12}+\Delta_{34}}\int {\cal D}\exp\left(-i\pi G_N\ell p^up^v|\sin(\phi-\phi')| -2\ell p^u \epsilon_{13} - 2\ell e^{-t/\ell}p^v\epsilon_{24}^* - t/\ell\right) ~. \nn
\end{align}
First performing the integrals over $\phi,\phi'$ and $p$ we find
\be
F(t) = \frac{4 \ell \epsilon _{24}^* }{\pi\sqrt{\frac{\pi ^2 q^2 G_N^2}{\epsilon _{13}^2}+4}}\exp\left(-\frac{t}{\ell}-2 \ell q \epsilon _{24}^* e^{-\frac{t}{\ell}} \right) \left(\pi -2 i \,\text{arcsinh}\left(\frac{\pi  q G_N}{2 \epsilon _{13}}\right)\right)
\ee
To perform the final integral over $q$, we focus on the regime $qG_N \lesssim 1$ such that we can ignore the arcsinh term.\footnote{As explained in \cite{Shenker:2014cwa}, this is an appropriate expansion because the integral is dominated by the regime $\delta \sim 1$ in de Sitter units. We can then take $qG_N\lesssim 1$ as long as $p\ell \gtrsim 1$.} In that case, we can perform the remaining integral analytically in terms of special functions, which gives the result
\be \label{eq:fullOTOC}
F(t) =g \left(\pi  H_0(2g)+2 {\cal F}(g^2)+2\log\left(-g\right)J_0(2 g)\right)~.
\ee
Here $H_n(z)$ is the Struve function, $J_n(z)$ the Bessel function of the first kind and we defined the function
\be
{\cal F}(z) = \lim_{a\to1}\partial_a\left(\frac{\,_0F_1(a,-z)}{\Gamma(a)}\right) ~,
\ee
as a limit of the confluent hypergeometric function. The argument $g(t)$ is defined by\footnote{The limit $\text{arg}(\epsilon_{13})\to0$ is to be taken from below.}
\be
g(t) = \text{sgn}(\text{arg}(\epsilon_{13}))\frac{2 \ell\epsilon_{13}\epsilon_{24}^*e^{-t/l}}{\pi G_N} ~.
\ee
To compare this result to the geodesic approximation\footnote{Strictly speaking, we can only compare at times earlier than the scrambling time, since the operators $W_R$ and $V_L$ are spacelike separated in that case and $\braket{W_RV_LV_RW_R}=\braket{V_LW_RV_RW_R}$.}, we need to send one of the $V$ operators to the $L$ patch. We can do this by taking $\epsilon_1 = - \pi \ell, \epsilon_3 = 0$ which sends $\epsilon_{13} \to - 2 i$. Next, we would like to send $\epsilon_{24}\to 0$, but in this limit the correlation function vanishes. As was explained in \cite{Shenker:2014cwa,Roberts:2014ifa} this is due to the high-energy nature of the $W$ operators, which we are trying to evaluate at the same point. This behaviour can be regulated by smearing the operators over a thermal length $\beta$ before sending $\epsilon_{24}\to0$. Instead of doing this explicitly, we will instead leave $\epsilon_{24}$ finite just as was done in \cite{Shenker:2014cwa,Roberts:2014ifa} and think of $\epsilon_{24} \sim {\cal O}(1)$. Explicitly, we will take $\epsilon_4=-\epsilon_2=\tau\ell > 0$, which sends $\epsilon_{24}\to2\sin\tau$. For these values of $\epsilon_{ij}$, the correlation function is shown in figure \ref{fig:OTOC}.
For early times we find that, just as in the geodesic approximation, the real part of the correlation function increases. For later times however, the correlation function decreases and goes to zero. This should be contrasted with the behaviour of the OTOC in the geodesic approximation \eqref{eq:lateOTOC} which oscillates. As mentioned before, this qualitatively different behaviour can likely be attributed to the different regime of mass that we are considering.

Because a geodesic crossing a shockwave with positive null energy in de Sitter space experiences a time advance, it becomes possible to send signals from the left patch $L$ to the right patch $R$. In this sense, de Sitter space shares similarities with traversable wormholes in Anti-de Sitter space \cite{Gao:2016bin,Maldacena:2017axo}, with the important difference that there is no need for a non-local coupling between the poles. To confirm traversability, we can consider the response of an operator $V_R(0)$ to a perturbation to the left static patch by an operator $e^{i\epsilon_LV_L(0)}$ once we include the particle $W_R(t)$ that creates a shockwave. For $W,V$ Hermitian operators this response is given by \cite{Maldacena:2017axo}
\begin{align}
&\braket{e^{-i\epsilon_LV_L(0)}W_R(t)V_R(0)W_R(t)e^{i\epsilon_L V_L(0)}} = \\
&\qquad \braket{W_R(t)V_R(0)W_R(t)} + 2\epsilon_L\,\im{(\braket{V_L(0)W_R(t)V_R(0)W_R(t)})} + {\cal O}(\epsilon_L^2) ~. \nonumber
\end{align}
An imaginary part of the OTOC
\be
F(t) = \braket{V_L(0)W_R(t)V_R(0)W_R(t)} ~,
\ee
therefore shows that a signal has been exchanged between the left and right static patch, because the expectation value of $V_R(0)$ in the shockwave background depends on the left perturbation $\epsilon_L$. This is precisely the correlator that has been plotted in figure \ref{fig:OTOC}, showing that the wormhole connecting the left and right static patch opens up due to the shockwave. 

\begin{figure}[h]
\centering
\includegraphics[scale=.8]{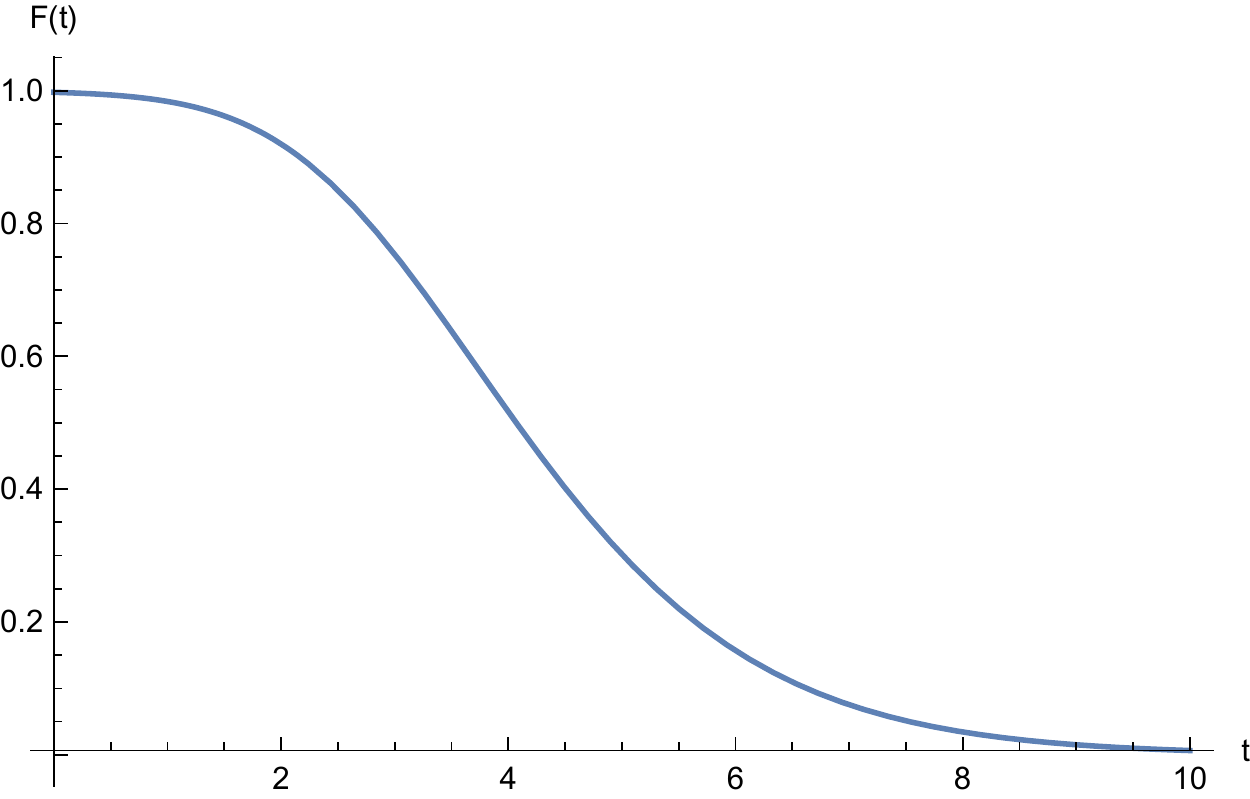}
\caption{The out-of-time-order correlator $\braket{V_R(0)W_R(t)V_R(0)W_R(t)}$ for $|g(t=0)|=10,\ell=1$.}
\label{fig:OTOC2}
\end{figure}

It is also interesting to consider OTOCs with operators inserted at different points. For example, we can also consider a purely single-sided correlator. Since both the $W$ and $V$ operators are inserted at the same point, we have to regulate this correlation function. This can be done by taking $\epsilon_4 = \epsilon_3 = -\epsilon_2=-\epsilon_1=\tau \ell > 0$. The resulting OTOC is displayed in \ref{fig:OTOC2}. 
Expanding for $|g(t)| \gg 1$ (and setting $2\sin\tau = 1$) we now find
\be \label{eq:expandedOTOC2}
F(t) = 1 - \left(\frac{G_N\pi}{2\ell} e^{t/\ell}\right)^2 + {\cal O}\left(\frac{G_N}{\ell}e^{t/\ell}\right)^4~.
\ee
We therefore see that at times $\ell \ll t \ll \ell \log(S_{dS}) $ this correlator decreases exponentially. Notice that the second term of \eqref{eq:expandedOTOC2} comes with a square, which is different than in the black hole case. Nonetheless, the timescale where the OTOC is affected by an order one amount is the same and given by the scrambling time $t_* = \frac{\beta}{2\pi}\log(S)$. The fact that the leading term in the OTOC proportional to the entropy goes as $1/S^2$ instead of $1/S$ (as in black holes) might be an important hint about the different microscopic structure of de Sitter space as compared to black holes. In any case, we find that the Lyapunov exponent of the purely single-sided OTOC is given by $\lambda_L=2\pi/\beta$. This shows that the de Sitter horizon space is a `fast scrambler' that saturates the chaos bound \cite{Maldacena:2015waa}.

\subsection{Stringy corrections}
Because of the large blueshift perturbations experience, the scattering process of perturbations with rest energy $E_0$ necessarily involves transplanckian energies when they are released with a time separation greater than $t = \ell \log(m_p/E_0)$, where $m_p$ is the Planck mass. For thermal quanta with rest energy $E_0\sim 1/\ell $ this is proportional to the scrambling time. As such, one can wonder about the validity of our computation of the OTOC.

For black holes in Anti-de Sitter space it turns out that such quantum gravity corrections are surprisingly mild \cite{Shenker:2014cwa}. The main corrections are due to the softer UV behaviour of string amplitudes, as the Eikonal phase grows with the center-of-mass energy $s$ as
\be
\delta \sim \sum_J G_N s^{J-1} ~.
\ee
Here, $J$ is the spin of the particles that contribute. In Einstein gravity, this is dominated by the graviton $(J=2)$ leading to a linear dependence on $s$.  In string theory on the other hand we need to sum over an entire tower of higher-spin states leading to a slower growth of \cite{Jahnke:2018off}
\be
\delta \sim G_N s ^{J_{\rm eff} -1} ~,
\ee
where
\be
J_{\rm eff} = 2 - d(d+1)\frac{\ell_s^2}{\ell^2} ~,
\ee
with $\ell$ the AdS length and $\ell_s$ the string length. This implies that chaos develops slower leading to a scrambling time of \cite{Shenker:2014cwa}
\be
t_* = \frac{\beta}{2\pi}\left(1+ \frac{d(d+1)}{4}\frac{\ell_s^2}{\ell^2} + \dots \right)\log(S) ~,
\ee
where the dots denote terms higher order in $\ell_s^2/\ell^2$.

In de Sitter space, we would like to make a similar argument. An additional complication however is that in de Sitter space there exists a bound on the mass of higher-spin states to fall into unitary representation of the isometry group. This bound, known as the Higuchi bound, is given by \cite{Higuchi1987}
\be
m^2\ell^2 \geq  (J-1)(d-4+J) ~.
\ee
As a consequence, for a linear Regge trajectory $m^2\ell_s^2=J$ the Higuchi bound is violated in three dimensions at spin  \cite{Noumi:2019ohm,Lust:2019lmq}
\be \label{eq:higuchiviolation}
J \gtrsim \frac{\ell^2}{\ell_s^2} ~.
\ee
If gravity is UV completed by the leading Regge trajectory in a weakly coupled regime, we need a sufficiently large number of higher-spin states at energies $m_s<E<\Lambda$, where $m_s$ is the string scale and $\Lambda$ the cutoff of the theory. This implies that the mass of the maximum spin state consistent with the Higuchi bound should be above the cutoff. Taking the cutoff to be the Planck scale $\Lambda=m_p$ this implies a bound on the Hubble parameter $H=1/\ell$.
\be
H \lesssim \frac{m_s^2}{m_p} ~.
\ee
If this bound is satisfied the UV behaviour of the scattering amplitude is softened and we expect the scrambling time to increase by including stringy effects. In that case, the scrambling time we derived in Einstein gravity should be viewed as a lower bound.

\section{Consequences for complementarity} \label{sec:complementarity}
 After studying chaos in de Sitter space, we now turn to discuss the consequences of our results for observer complementarity. In the context of black holes, observer complementarity \cite{Susskind:1993if} (see also \cite{Hooft1985,Hooft1990}) suggests that the infalling and asymptotic observer might have a completely different, but complementary experience. While the asymptotic observer sees the infalling observer blueshift and reach a Planckian temperature, the infalling observer would report that nothing dramatic happened when she crossed the horizon. These two different perspectives are `complementary', because the two observers are never able to meet up again and report on their experience.

For black holes, this idea can be challenged by considering a thought experiment \cite{Hayden:2007cs,Sekino:2008he} in which the infalling observer (Alice) carries a qubit and immediately sends it parallel to the future horizon after crossing it, see figure 1 of \cite{Sekino:2008he}. At the moment that Alice has crossed the horizon, the black hole contains this qubit of information and will eventually reemit it in the form of Hawking radiation. The asymptotic observer (Bob) waits until he has collected enough radiation to decode Alice's qubit and then jumps into the black hole after her. If the time it takes for Bob to decode Alice's qubit is short enough, he will be able to receive Alice's qubit directly from her before it is destroyed by the singularity. Thus, he will observe the same qubit twice in violation of quantum no-cloning and complementarity.

The resolution to this paradox comes from the fact that the minimum amount of time it takes for Bob to decode Alice's bit is the scrambling time $t_*=\frac{\beta}{2\pi}\log(S)$ \cite{Hayden:2007cs,Sekino:2008he}, which is just long enough to prevent an observable violation of no-cloning. Still, it leads to the perhaps unsatisfactory point of view that the message itself is cloned, although there is no observer to witness it. Traversable wormholes in Anti-de Sitter space \cite{Gao:2016bin,Maldacena:2017axo} have put a new perspective on this. If Bob collects a large amount of Hawking radiation and collapses it to a black hole that is maximally entangled with Alice's black hole, he has created the thermofield double state. Then, as explained in \cite{Maldacena:2017axo} the action of Bob decoding Alice's qubit essentially corresponds to the situation where the qubit traverses the wormhole and moves from one boundary to the other. At all times, there is just one copy of Alice's qubit in the system.

Now let's turn our attention to de Sitter space. In de Sitter space, it is reasonable to expect that a similar notion of complementarity should exist between a static observer (Bob) and a freely falling observer (Alice) \cite{Bousso:2000nf,Banks:2001yp,Goheer:2002vf,Dyson:2002pf,Parikh:2002py,Parikh:2008iu}. However, as was highlighted in \cite{Parikh:2008iu} in de Sitter space it is not possible for Bob to decode even one bit of information from the Hawking radiation.\footnote{The situation is different for inflationary spacetimes in which the exponential expansion ends (locally) \cite{Danielsson:2002td,ArkaniHamed:2007ky,Dubovsky:2008rf,Nomura:2011dt,Huang:2012eu}. In that case, information about the inflationary past of the universe can be retrieved at late times.} In four dimensions, this is essentially due to the finite volume of the static patch, which causes Bob's patch to collapse to a black hole when he tries to do so.\footnote{The amount of quanta that need to be collected is $S_{dS}/2$ \cite{Page:1993df}, but the maximum amount of entropy that can be stored in a single static patch is given by the Nariai black hole which in four dimensions has an entropy of $S_{dS}/3$.}

\begin{figure}[h]
\centering
\includegraphics[scale=.5]{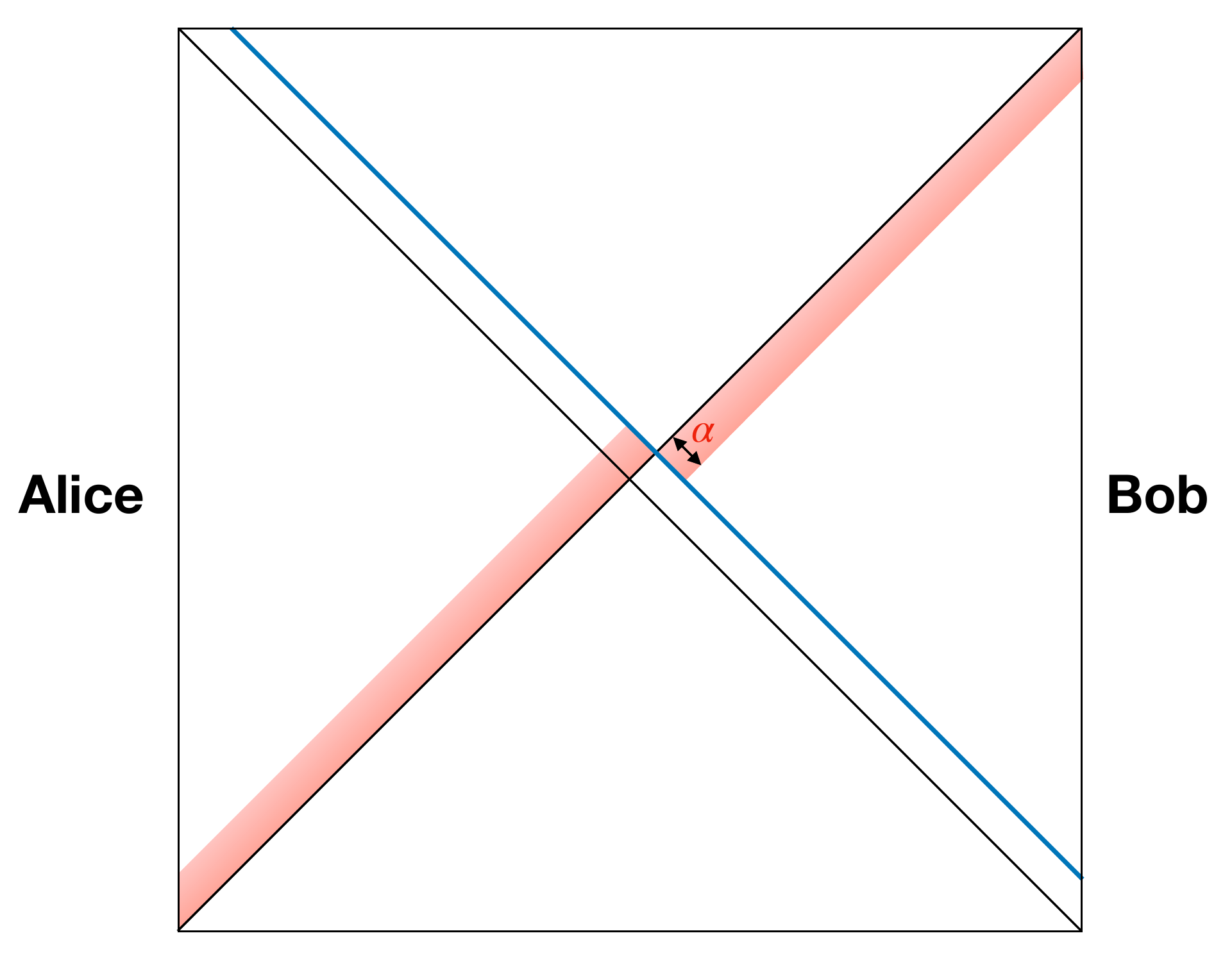}
\caption{Bob is sitting at the South pole of de Sitter space and creates a positive energy shockwave (blue). As a result, Alice can send a message (red) to Bob at the North pole.}
\label{fig:ABindS}
\end{figure}

But the situation is different when we allow for perturbations to de Sitter space. As we discussed in section \ref{sec:dSbasics}, positive energy perturbations that are released from the South pole lead after a scrambling time to a geometry in which the left and right static patch of the global de Sitter Penrose diagram are causally connected. If Alice sits at the North pole in such a geometry, she can send a message to Bob at the South pole, see figure \ref{fig:ABindS}.
However, it should not be possible for Alice to send arbitrary large amounts of information to Bob. Complementarity suggests that Bob should have access to $S_{dS}$ bits of information at most, so what prevents Alice from sending more information? The proper time that the wormhole is open is given by
\be
\Delta\tau = \frac{2\ell^2}{\ell^2-uv}\sqrt{\Delta u \Delta v} ~.
\ee
Close to the horizon ($v=0$) this leads to $\Delta\tau = 2\alpha$, where $\alpha$ is related to the energy of the particle generating the shockwave.
\be
\alpha = \pi G_N\ell p^u ~.
\ee
Here, $p^u$ is the energy of the shockwave. If the shockwave is generated by a particle with a thermal energy $E_0 = \beta^{-1}$, in the restframe of the shockwave the wormhole is only open for a Planckian proper time: $\Delta\tau = G_N$. Nonetheless, as stressed in \cite{Freivogel:2019whb} this does not imply that Alice needs to finetune the timing of her message to make sure it passes through the wormhole and reaches Bob. From her perspective, the wormhole is open exponentially longer due to a large time delay between a clock at the horizon and Alice's clock. There are two conditions that need to be satisfied for Alice's message to reach Bob \cite{Freivogel:2019whb}. First of all, we would like the energy of the message to be small enough such that it does not backreact strongly on the background geometry. This translates to the condition
\be \label{eq:probecond}
p^{\rm tot} < \frac1{G_N} ~,
\ee
where $p^{\rm tot}$ is the total energy of Alice's message.  At the same time, for the message to fit through the wormhole its wavelength should be sufficiently blueshifted. Denoting the energy of a single bit $N$ of Alice's message by $p^v$ such that $p^{\rm tot}= N p^v$, this amounts to satisfying $p^v > 1/\alpha$. Using the rest frame energy of the shockwave, this becomes
\be \label{eq:blueshiftcond}
p^v > \frac 2{G_N} ~.
\ee
Now, combining \eqref{eq:probecond} and \eqref{eq:blueshiftcond} we find that the number of bits Alice can successfully send is bounded by
\be \label{eq:singlefieldbound}
N \lesssim 1 ~.
\ee
So Alice can only send an ${\cal O}(1)$ number of bits! The prospects for information exchange become better if Bob uses a large number $K\gg1$ of light species to create the positive energy shockwave. In this case, the amount of energy and therefore the time that the wormhole is open is enhanced by a factor of $K$ \cite{Freivogel:2019whb}. At the same time, the probe condition \eqref{eq:probecond} remains unchanged, such that the total number of bits is now bounded by
\be
N \lesssim K ~.
\ee 
Of course, $K$ cannot be arbitrary large, because a large number of species changes the cutoff to $\ell_{\rm UV} \gtrsim G_NK $. For a semi-classical description we have to impose $\ell_{\rm UV} \ll \ell$ and the maximum amount of bits becomes bounded by
\be \label{eq:multifieldbound}
N \lesssim \frac{\ell}{G_N} \simeq S_{dS} ~.
\ee
The same result can be obtained by viewing the effect of introducing $K$ light species as a renormalization of $G_N$, while keeping the cutoff fixed.\footnote{We thank Antonio Rotundo for helpful discussions regarding this point.} In any case, this bound agrees with the intuition that Bob should have access to $S_{dS}$ bits of information at most. At the moment Alice tries to send more, her message either does not fit through the wormhole, or backreacts. Similar observations have been made in \cite{Leblond:2002ns}.

\subsection{Implications for inflation}
Although the main results of this work were derived in the context of (perturbations to) pure de Sitter space, we will now briefly speculate how these results might have consequences for inflationary scenarios in which the Hubble parameter slowly evolves with time.

Despite the different global structure of inflationary spacetimes, which only cover half the Penrose diagram, we expect that at least some of the features we observed in de Sitter space carry over to inflation. In particular, it is well known that an observer in quasi-de Sitter space is surrounded by a horizon that can be attributed thermodynamic properties just as in pure de Sitter space \cite{Frolov:2002va}. One of the consequences of this is that we still expect that the time it takes for a perturbation to the horizon to be indistinguishable from its thermal atmosphere is given by $t =\ell \log\left(S\right)$, i.e. the scrambling time. This is of interest, because the scrambling time recently made its appearance in cosmology via the papers \cite{Bedroya:2019snp,Bedroya:2019tba}. There, it was suggested that an inflationary fluctuation should always have a wavelength longer than a Planck length which puts a bound on the number of e-folds of inflation, given by $N_e < \log(m_p/H)$. We should stress that we don't yet have a compelling argument in favor for this conjecture, but if the fundamental timescale governing it is the scrambling time (as hinted upon in \cite{Bedroya:2019tba}) this bound should instead read\footnote{The conjecture of \cite{Bedroya:2019snp,Bedroya:2019tba} is phrased in terms of the `inflationary' time coordinate in the metric \eqref{eq:FRWmetric}, whereas the scrambling time is given in terms of the static time coordinate in the metric \eqref{eq:staticmetric}. However, at the center of the static patch these two different notions of time coincide.} $N_{e} < \log(S_{dS})$, where $S_{dS}$ is the de Sitter entropy. This seems like a minor modification, but in four dimensions the de Sitter entropy scales as $S_{dS} \sim m_p^2/H^2$ effectively doubling the number of e-folds allowed.

Furthermore, we also expect that the observation that a perturbation to pure de Sitter space leads to a traversable wormhole can be given an interpretation in quasi-de Sitter space. To do so, we consider two observers in different Hubble patches that have been generated during inflation. We then assume that the quantum state of these two patches is (approximately) described by a maximally entangled state. This is a strong assumption that we intend to explore further in future work, but one that seems natural: it is well known that in order to have a smooth horizon, (maximum) entanglement between regions separated by a horizon is key. If this assumption holds, these two Hubble patches are effectively described by two static patches, just as in pure de Sitter space. A shockwave that connects the two patches can now arise as follows. During inflation there is a positive energy flux out of the horizon due to a slowly decreasing Hubble parameter. As long as the Hubble parameter is evolving slowly, the amount of energy flux is given by the thermodynamic relation $dE = TdS$ \cite{Frolov:2002va}. In four dimensions, we can write this as
\be
\dot E = \frac{\epsilon}{G_N} ~,
\ee
where $\epsilon = -\dot H/H^2$ is the first slow-roll parameter. Here the dot denotes a time derivative with respect to the cosmological time $t$, as measured in the metric
\be \label{eq:FRWmetric}
ds^2 = -dt^2 + e^{2H(t)}d\vec x_3^2 ~,
\ee
Taking $\epsilon \ll 1$ to be constant, the energy flux that leaves the horizon in a Hubble time $t=1/H$ is given by
\be
E = \frac{\epsilon}{G_N H} ~.
\ee
This is appropriately described as a positive energy shockwave when the energy in the rest frame of the flux is given by $E \geq H$, leading to
\be
\epsilon \geq \frac{H^2}{8\pi m_p^2} ~.
\ee
At the same time, for inflation to be semi-classical we need to impose $H/m_p \ll 1$. This also ensures that the classical growth of the horizon is negligible during a Hubble time. Combining these bounds we find that the energy flux leaving the horizon during inflation can be appropriately described by a shockwave in the regime
\be
\frac{H^2}{8\pi m_p^2}\leq \epsilon \ll 1~.
\ee                                                                                                                                                              
The lower bound on $\epsilon$ also prevents a transition to (slow-roll) eternal inflation. In this regime of parameters it becomes possible for information to enter a Hubble patch from a previously causally disconnected patch after the positive energy has sufficiently blueshifted to form a shockwave. This happens after $N_e = \log(S_{dS})$ e-folds of inflation. If only a single field contributes to the positive energy of the shockwave, the bound \eqref{eq:singlefieldbound} applies and at the moment that more than ${\cal O}(1)$ bits of information enter, this information flow will lead to backreaction. If, on the other hand, a large number of light fields contributes \eqref{eq:multifieldbound} applies and we can ignore backreaction until ${\cal O}(S_{dS})$ bits have entered. If we assume that there is roughly one bit of information that enters per e-fold\footnote{If we view the de Sitter horizon as encoding one bit per Planck area, inflation typically generates one bit of information per e-fold \cite{ArkaniHamed:2007ky}.}, this implies that backreaction effects can safely be ignored for $N_e \lesssim \log(S_{dS})$ when a single light field contributes or $N_e \lesssim S_{dS}$ when the maximum allowed number of fields contribute. We should stress however that we are not suggesting that inflation terminates after this time. Our results only give a criterium when backreaction effects become important if information enters the Hubble patch by crossing the shockwave. If this does not happen, such a bound does not apply.

\section{Discussion} \label{sec:discussion}
In this paper, we studied chaos in de Sitter space by computing several out-of-time-order correlators (OTOCs) of scalar operators inserted at the center of the static patch. One of our main results is the observation that the purely single-sided OTOC consisting of four conformally coupled scalar fields exhibits maximal chaos. It decreases exponentially with a Lyapunov exponent that saturates the chaos bound $\lambda_L \leq 2\pi/\beta$. An interesting difference between black hole and de Sitter chaos is the fact that the leading term in the de Sitter OTOC is proportional to $1/S_{dS}^2$ and it would be satisfactory to better understand the underlying reason for this.

We should mention that our conclusion that the de Sitter horizon is maximally chaotic is different than \cite{Anninos:2018svg}. In that paper, the OTOC these authors calculated did not show Lyapunov behaviour, but exhibited oscillations. It should be kept in mind however that their setup is slightly different. Firstly, the OTOC that \cite{Anninos:2018svg} considered does not correspond to the purely single-sided configuration that we found displays maximal chaos. Secondly, we focussed on an OTOC with conformally coupled field, whereas \cite{Anninos:2018svg} considered massless perturbations. Massless fields might behave qualitatively different, since there is no vacuum state for a massless scalar field in de Sitter space that is invariant under the full isometry group \cite{Allen:1987tz}.

We also computed an OTOC where one of the operators is moved to the other pole of de Sitter space and found that it behaves differently: it initially increases and develops an imaginary part. We explained that this behaviour can be attributed to the fact that shockwaves that satisfy the null energy condition in de Sitter space bring opposite poles into causal contact, making the wormhole connecting the left and right static patch traversable. We discussed our results in the context of de Sitter complementarity and found that it is possible to send at most $S_{dS}$ bits of information through the wormhole. Sending more information than this leads to backreaction.

These results might have implications for inflation. Since in an inflationary phase positive energy is leaving a Hubble patch, this energy can be appropriately described by a shockwave in a certain regime of parameters that we specified. When this happens, information from a previously causally disconnected part of spacetime can enter the Hubble patch. If this amount of information becomes too large, backreaction cannot be ignored. Although this does not directly put a bound on the number of e-folds of inflation, such as in \cite{Bedroya:2019snp,Bedroya:2019tba}, it does clarify the meaning of the scrambling time in an inflationary spacetime. 

In future work, it would be interesting to consider OTOCs not only for conformally coupled fields but for arbitrary masses. Since the structure of the Wightman function is now much more complicated it might not be possible to do this analytically and one would have to resort to numerics. Related to this, using the recent developments in the cosmological bootstrap \cite{Arkani-Hamed:2018kmz,Baumann:2019oyu,Isono:2018rrb,Isono:2019ihz,Isono:2019wex} one might be able to directly write down four-point functions for fields with arbitrary masses. If it can then be confirmed that after an appropriate analytical continuation the OTOC also displays maximal chaos for very light fields (such as the inflaton), one can study the implications of maximal chaos on inflation. We hope to come back to some of these questions in a future correspondence.

\section*{Acknowledgments}
We thank Daniel Chung, Rob Myers, Antonio Rotundo and Jan Pieter van der Schaar for helpful discussions. This work is supported in part by the DOE under grant DE-SC0017647, the National Science Foundation under Grant No. NSF PHY-1748958, and the Kellett Award of the University of Wisconsin. GS gratefully acknowledges the hospitality of the Kavli Institute for Theoretical Physics during the final stage of this work.

%%%%%%%%%%%%%%%%%%%%%%

\appendix

\section{Derivation of the shockwave geometry} \label{app:shockwave}
Here we derive the shockwave geometry created by a particle that is released from the center of the static patch of de Sitter space. The approach we will take is adapted from \cite{Hotta:1992qy}. We start with the metric for a point particle in de Sitter space and perform a boost to generate the shockwave geometry. The three-dimensional Schwarzschild-de Sitter solution in static coordinates is given by \cite{Spradlin:2001pw}
\be
ds^2 = -\left(1-8G_Nm-\frac{r^2}{\ell^2}\right)dt^2 + \left(1-8G_Nm-\frac{r^2}{\ell^2}\right)^{-1}dr^2+r^2d\phi^2 ~.
\ee
This metric has a single horizon at $r=\sqrt{\ell^2-8G_N\ell^2m}$ and describes a point particle at the origin of the static patch. Expanding for small $G_Nm$ and writing the metric in embedding coordinates, we find
\be
ds^2 = ds^2_{0} + \frac{8G_N\ell^2m}{(X_0^2-X_3^2)^2}\left(\left(X_0 dX_3 - X_3 dX_0\right)^2 + \ell^2\frac{\left(X_0dX_0 - X_3dX_3\right)^2}{\ell^2+X_0^2-X_3^2}\right) ~.
\ee
Here $ds_0^2$ is the unperturbed de Sitter metric. We now perform a boost along the $X_1$ direction, which sends
\begin{align}
X_0 &\to \frac1{\sqrt{1-\beta^2}}\left(X_0 -\beta X_1\right) ~, \\
X_1 &\to \frac1{\sqrt{1-\beta^2}}\left(X_1 -\beta X_0\right) ~, \nn\\
m &\to p\sqrt{1-\beta^2} ~. \nn
\end{align}
Taking the ultrarelativistic limit $\beta\to 1$ the perturbation vanishes everywhere except at $-X_0+X_1=0$. Writing the metric in terms of $x = \frac{X_0-\beta X_1}{\sqrt{1-\beta^2}}$,  we find
\be
ds^2 = ds_0^2 + \frac{8G_Np\ell^2\sqrt{1-\beta^2}}{(x^2-X_3^2)^2}\left((xdX_3-X_3dx)^2+\ell^2\frac{(xdx-X_3dX_3)^2}{\ell^2+x^2-X_3^2}\right)
\ee
We now make use of the following representation of the limit $\beta\to 1$ \cite{Hotta:1992qy}.
\be
\lim_{\beta\to1}\frac1{\sqrt{1-\beta^2}} f\left(\frac{(X_0-\beta X_1)^2}{1-\beta^2}\right) = \delta(X_0-X_1)\int_{-\infty}^{+\infty}dx\,f(x^2) ~.
\ee
Evaluating the integral we then find the following metric.
\be
ds^2 = ds^2_0 - 8\pi G_N p|X_2|\,\delta(X_0-X_1)(dX_0-dX_1)^2 ~,
\ee
where we used $-X_0^2+X_1^2+X_2^2+X_3^2=\ell^2$.

We can write this metric in a more familiar form by picking coordinates in which de Sitter space has flat spatial slices.
\begin{align}
X^0 &= \frac{\ell^2-\eta^2+(\rho\cos\phi-\ell)^2+\rho^2\sin^2\phi}{2\eta} ~, \\
X^1 &= \frac{\ell}{\eta}(\rho\cos\phi-\ell) ~, \nn \\
X^2 &= \frac{\ell}{\eta}\rho\sin\phi ~, \nn \\
X^3 &= \frac{\ell^2+\eta^2-(\rho\cos\phi-\ell)^2-\rho^2\sin^2\phi}{2\eta} ~. \nn 
\end{align}
In these coordinates the metric becomes
\be
ds^2 =  \frac{\ell^2}{\eta^2}\left(-d\eta^2 + d\rho^2 + \rho^2d\phi^2\right) - 16\pi G_N p\ell\, |\sin\phi|\delta\left(\frac{\eta^2-\rho^2}{\eta}\right)\left(d\eta+d\rho\right)^2 ~.
\ee
Transforming to global coordinates $(u,v)$ (defined in \eqref{eq:KruskalEmbed}) we obtain
\be
ds^2 =  \frac{4\ell^4}{(\ell^2-uv)^2}(-dudv)-8\pi G_Np\ell{|\sin\phi|}\delta(u)du^2 + \ell^2\left(\frac{\ell^2+uv}{\ell^2-uv}\right)^2d\phi^2 ~.
\ee
This metric describes a shockwave travelling along the $u=0$ horizon and is a solution to Einstein's equations with a stress tensor given by
\be
T_{uu} = \frac{p}{2\ell}\delta(u)\delta(\sin\phi) ~,
\ee
We can also consider a shockwave travelling along the $v=0$ horizon, simply by interchanging $u\leftrightarrow v$.

\bibliographystyle{utphys}\bibliography{refs}

\end{document}